\documentclass[5p]{elsarticle}

\usepackage{lineno,hyperref}
\modulolinenumbers[5]

\usepackage[utf8x]{inputenc}
\usepackage{float}
\usepackage{amsmath}
\usepackage{amssymb}
\usepackage{graphicx}
\usepackage{color}

\journal{arxiv.org}

\newcommand{\rev}{``}

\begin{document}

\begin{frontmatter}

\title{Theory and phenomenology of relativistic corrections to the Heisenberg principle}


\author{Giovanni Amelino-Camelia}
\address{Dipartimento di Fisica Ettore Pancini, Università di Napoli Federico II, and INFN, Sezione di Napoli,\\ Complesso Univ. Monte S. Angelo, I-80126 Napoli, Italy}

\author{Valerio Astuti}
\ead{valerio.astuti@gmail.com}
\address{Dipartimento di Fisica, Università di Roma “La Sapienza”, P.le A. Moro 2, 00185 Roma, Italy\vspace{-0.3cm}}

\begin{abstract}
The Heisenberg position-momentum uncertainty principle shares with the equivalence principle the role of main
pillar of our current description of nature. However, in its original formulation it is inconsistent with special relativity, and in nearly a century of investigation not much progress has been made
toward a satisfactory reformulation. Some partial insight has been gained in the ultra-high-velocity regime  but a full description is still missing and in particular we have no clue about the intermediate regime of particles whose speeds are much smaller than the speed of light but still high enough for tangible departures from the Heisenberg formulation to be present.
As we stress here, that intermediate regime is also our best chance for testing experimentally our understanding of the implications of special relativity for the uncertainty principle.
We here introduce a new approach
to these challenges, based mainly on the observation that the only operative notion of position of a particle at a given time involves the crossing of the worldline of that particle with the worldline of
a test particle.  We find that the worldline-crossing perspective opens a path toward
a special-relativistic version of the uncertainty principle, which indeed could be tested experimentally.
\end{abstract}


\end{frontmatter}


Arguably the equivalence principle and Heisenberg's uncertainty principle are the two most
important aspects of our current description of nature, since they are the primary principles on
which general relativity and quantum mechanics are built. Any progress in reaching a deeper understanding
of the
equivalence principle and of the uncertainty principle might also be crucial for investigating
the conceptual incompatibility between
general relativity and quantum mechanics, which is the focus of a large research effort
(see, e.g., \cite{amelino2013quantum, carlip2001quantum} and references therein).
We are here concerned with a residual gray area in the understanding of the uncertainty
principle between position and momentum: its original formulation by Heisenberg neglects special-relativitic effects
and after nearly a century still not much progress has been made
on reformulating it consistently with special relativity.
The interface between special relativity and quantum mechanics is described by
special-relativistic quantum field theory, but the faith of the uncertainty principle remains unclear in the transition from the galilean-relativistic formulation of quantum mechanics to
special-relativistic quantum field theory.
The most popular description of the uncertainty
principle is in terms of a relation between the uncertainty in position predictions and the uncertainty in momentum predictions, but the position of a particle is not a good observable
of special-relativistic quantum field theory (it only emerges as a good observable in the galilean-relativistic limit) and therefore also its uncertainty becomes meaningless.

\begin{figure}[h!]
\includegraphics[scale=0.85]{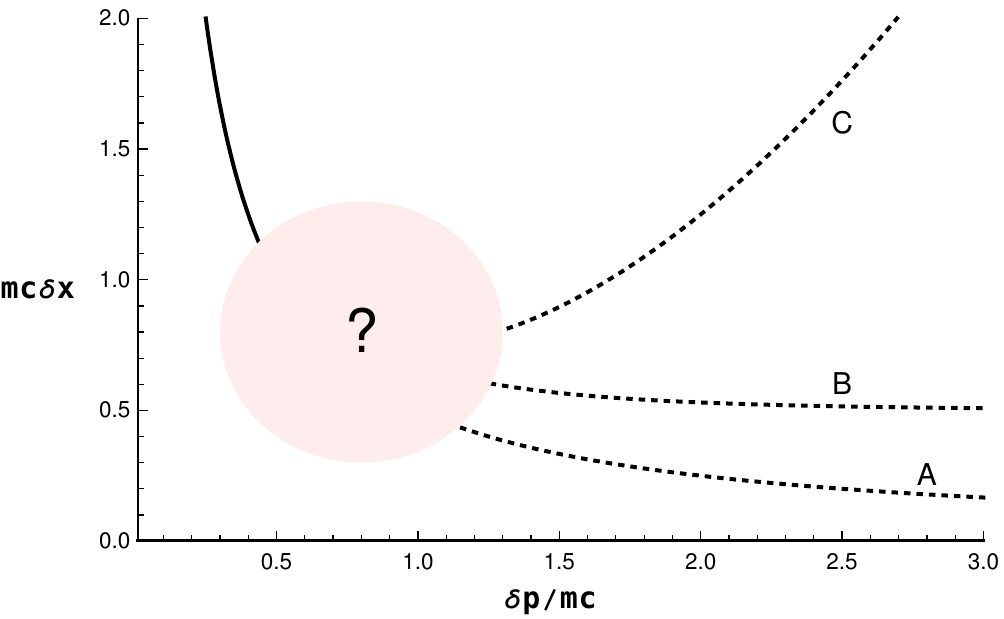}
\caption{\scriptsize{How accurately can we determine the position of a particle at rest for a given uncertainty in its momentum? This is operatively a well-defined question in all regimes of physics, even though our current formalisms in some cases are unable to handle it. For low momentum uncertainty we are confident that the Heisenberg principle governs the minimum position uncertainty. For high momentum uncertainty we do not expect the Heinseberg principle (\textbf{A}) to hold. The heuristic argument of \cite{landau1931erweiterung} suggests (\textbf{B}) that the minimum position uncertainty should reach an horizontal asymptote at about $\delta x \approx \frac{\hbar}{m c}$, but it is nothing more than a tentative argument and for example it remains possible that the minimum position uncertainty would at some point start growing (\textbf{C}). While these scenarios for the faith of the uncertainty principle at very high momentum uncertainty are conceptually intriguing, we here argue that the more urgent questions reside in the region highlighted in figure with a question mark, the regime where momentum uncertainty is still small but special-relativistic corrections could be appreciated, which is also the regime best suited for testing experimentally our understanding of the interplay between special relativity and the uncertainty principle.}}
\label{fig:regimes}
\end{figure}

This has led to the popular viewpoint that there is no middle ground: the uncertainty principle between position and momentum goes abruptly from
being a cornerstone of quantum mechanics in the galilean regime to being meaningless as soon as some
special-relativistic effects are taken into account.
We feel (like others, see e.g. Refs.\cite{landau1931erweiterung,newton1949localized,freidel2007relativistic} and references therein)
that the faith of the uncertainty principle in this transition cannot be so abrupt and that a lot of
insight on the fundamental workings of nature may reside in learning how to properly describe
the transition (see figure \ref{fig:regimes}).
Think for example of varying the precision of the measuring devices used
in an experiment involving only particles of very low speeds: for low precision of the measuring devices
special-relativistic effects cannot be appreciated and Heisenberg's uncertainty principle would surely reign supreme
upon the measurement outcomes. Should we then expect that when the precision of the devices is increased just up to the threshold for appreciating the first special-relativistic effects,
 Heisenberg's uncertainty principle would become useless for the description of the measurement outcomes?
 We feel that this is unlikely, but most importantly we feel that a scientific answer to this question
 is needed, a challenge for theories at first, but ultimately a question to be settled experimentally. The only plausible scenario is that when special-relativistic effects are just barely appreciable the Heisenberg principle should still be at center stage, only reflecting those special-relativistic effects through some extra terms (\rev corrections") to its formulation.
 
To our knowledge the first study to provide insights relevant for this question dates back
to the 1930s: in Ref.\cite{landau1931erweiterung} Landau and Peierls
used the criterion that any viable position measurement procedure
should not produce an energy fluctuation capable of creating new particles, thereby proposing
a lower bound on the uncertainty of the position
of a particle at rest given approximately by the Compton wavelength.
Other approaches \cite{newton1949localized,freidel2007relativistic} have relied on observables in the single-particle special-relativistic Hilbert space which have some (but not all)
of the properties of the spatial coordinate of the particle, but the outcome of that strategy of investigation remains uncertain, particularly because of some challenges related to causality violations \cite{wigner1983interpretation, teller1994interpretive}. 
Rather than looking for candidate operators on the single-particle special-relativistic Hilbert space
  we here propose that the  special-relativistic deformations of the Heisenberg description
 should be introduced at the level of the properties of the probability distribution obtained from the wave function. Our conceptual motivation stems from the
  intuition that a position measurement never truly establishes a single-particle property:
  it has to do with the result of a collision between two particles,
  the particle viewed as \rev system"
  in the measurement procedure and a test particle (the space-time point one is localizing in the classical limit is the point of crossing of the two worldlines).
For what concerns the description of the particle localization in terms of a probability distribution defined in space-time, our approach is similar to the one introduced in \cite{marolf2002relativistic, reisenberger2002spacetime}, which, however, do not describe the localization procedure in terms of actual interactions between system particles and probe particles (a crucial aspect of our approach).
As we shall show, the properties of the probability distribution obtained from the wave function can naturally make room for this collision perspective: we set it up as dependent on the properties of
both particles, though ultimately the properties of the test particle (which is \rev prepared" by the experimenter) become only implicit.
Our new postulate for the  probability density of finding a particle in a given position is actually derived from a schematic description of the position-measuring process, in which a \rev probe" particle is set up to interact with the  \rev system" particle whose position is of interest.
The probability of interacting with the system particle as a function of the probe position is clearly proportional to the probability of these particles \emph{being} in the same place\footnote{We shall not dwell much on that but the observable usually described as \rev position of a particle" actually is the position at which that particle interacts with the probe. There is no need to define the notion of \rev position of a particle" when that position is not measured
(if it is not measured it is not of concern for physics).} (assuming the interaction can be described as a contact interaction).
We start from the probability amplitude of the interaction between our ``system" particle with mass $m$ and a probe particle with mass $M\gg m$, as derived by standard quantum field theory:
\begin{multline}
\langle \psi_{e_{i}} \phi_{p_{i}}|e_{o}p_{o}\rangle=\int\frac{d^{3}e_{i}d^{3}p_{i}}{E_{e_{i}}E_{p_{i}}}\overline{\psi}_{e_{i}}\overline{\phi}_{p_{i}} \times \\ \times \delta^{4}\left(e_{i}+p_{i}-e_{o}-p_{o}\right) \langle e_{i}p_{i}|T|e_{o}p_{o}\rangle
\end{multline}
where $e_i$ and $p_i$ are the momenta of the incoming system and probe particles, $e_o$ and $p_o$ the momenta of the outgoing particles, and $T$ the transition matrix of the process. We are assuming the in-particles to be in the states $\psi_{e_i}$ and $\phi_{p_i}$, and the out-states to be plane waves.
We focus on elastic scattering because we will use a first order approximation of the theory in which the only non-trivial out-state consists of the same kind of particles of the in-state. 
It is important to note that this level of approximation has an operative counterpart which was not considered in the heuristic argument made
in Ref.\cite{landau1931erweiterung} in the 1930s: 
rather than focusing exclusively on what might be the smallest uncertainty achievable in a position measurement (needing large momentum uncertainty), we are here mainly concerned with the first-order special-relativistic corrections applicable when momentum uncertainty is still rather small.
Some of the implications of special relativity that are found \cite{landau1931erweiterung} when the
momentum uncertainty is large do not apply to the regime of relatively low momentum uncertainty where
we plan to predict (and eventually test experimentally) the uncertainty principle.

In the first order approximation we can substitute the transmission matrix $\langle e_{i}p_{i}|T|e_{o}p_{o}\rangle$ with the constant value $\lambda$ (the coupling constant of our theory). We will be interested only in ratios between probabilities, so this approximation can be made exact considering smaller and smaller values of the coupling constant \cite{itzykson2012quantum}.
To obtain the probability for the two in-particles to interact we have to consider the squared modulus of the amplitude and sum over all possible out-states.
This gives us:
\begin{multline}
P\left[\psi, \phi \right] \propto \int\frac{d^{3}e_{i}d^{3}p_{i}d^{3}e'_{i}d^{3}p'_{i}}{E_{e_{i}}E_{p_{i}}E_{e'_{i}}E_{p'_{i}}}\overline{\psi}_{e_{i}}\overline{\phi}_{p_{i}}\psi_{e'_{i}}\phi_{p'_{i}} \times \\ \times \delta^{4}\left(e'_{i}+p'_{i}-e_{i}-p_{i}\right) I\left(e_i, p_i \right)
\end{multline}
with
\begin{equation}
I\left(e_i, p_i \right) = \int\frac{d^{3}e_{o}d^{3}p_{o}}{E_{e_{o}}E_{p_{o}}}\delta^{4}\left(e_{i}+p_{i}-e_{o}-p_{o}\right)
\end{equation}
The function $I$ is an invariant function of the incoming momenta, so it can be evaluated in the center-of-mass reference frame. It can depend only on the scalar product $e_i \cdot p_i$, which in turn is a function of the center-of-mass momentum $p^*$ and the two masses $m$ and $M$. We focus on probe particles of very high mass and momentum, while the system particle is approximately at rest, thus the center-of-mass momentum $p^*$ is very large and we have:
\begin{equation}
\lim_{\left|p^* \right| \to \infty} I\left(e_i, p_i \right) = \tilde{I}\left(M^2+m^2 \right) 
\end{equation}
where $\tilde{I}$ is a dimensionless function of its argument which  can be factored out from the probability, obtaining:
\begin{eqnarray}
P\left[\psi, \phi \right] &\propto & \int\frac{d^{3}e_{i}d^{3}p_{i}d^{3}e'_{i}d^{3}p'_{i}}{E_{e_{i}}E_{p_{i}}E_{e'_{i}}E_{p'_{i}}}\overline{\psi}_{e_{i}}\overline{\phi}_{p_{i}}\psi_{e'_{i}}\phi_{p'_{i}} \times \nonumber \\
&& \times\, \delta^{4}\left(e'_{i}+p'_{i}-e_{i}-p_{i}\right)  \propto \nonumber \\
&\propto & \int d^4x \left| \psi(x) \right|^2 \left| \phi(x) \right|^2 \nonumber
\end{eqnarray}
with\footnote{Throughout the derivation we are considering units in which $\hbar = 1$.}
\begin{equation}
\psi(x) = \int \frac{d^3 p}{E_p} \psi_p \exp{i \left(\frac{E_p}{c} x_0 - \vec{p}\cdot \vec{x} \right)}
\end{equation}
Selecting for the probe particle $\phi$ a well localized state\footnote{Here \rev well-localized" means much better localized than the state of the system/target particle; this is not a problematic requirement, since one can exploit the fact that the probe could have arbitrarily large mass.}, $\left| \phi(x) \right|^2 \approx \int d\tau \delta^4\left(x - x(\tau) \right)$, we obtain for the probability:
\begin{equation}
P\left[\psi, \phi \right] \approx N \int d\tau \left| \psi(x(\tau)) \right|^2
\end{equation}
with $N$ a normalization constant.
This is an incoherent sum of the probability density of interacting at every instant of (proper) time $\tau$, thus we obtain for the probability density for the particle to be in position $x(\tau)$ at the time $\tau$ (on the world-line of the probe):
\begin{equation}
\rho\left(\tau\right) \propto \left| \psi(x(\tau)) \right|^2
\end{equation}
This is obviously also proportional to the probability density $\rho\left(t\right)$, $x(t)$ being the probe worldline expressed as a function of time $t$:
\begin{equation}
P\left[\psi, \phi \right] \approx N' \int dt\, \left| \psi(x(t)) \right|^2
\end{equation}
\begin{equation}
x\left(t \right)=\left(\begin{array}{c}
c t_0 + c t\\
\vec{x}_0 + \vec{v} t
\end{array}\right)
\end{equation}
This probability density is derived from the probability of interacting with a single probe, so it cannot be used as a probability density over all of space at a given instant, but only as a probability density of interacting at a given space-time point on the probe world-line. 

In this first exploratory study we focus on the simplified context of only one spatial dimension in quantifying the corrections to the Heisenberg principle that follow from our probability density.
We have that the probability of interaction between the probe and the target particle at time $t$ is proportional to $\left|\psi\left(x\left(t \right)\right)\right|^2$, with:
\begin{equation}
    \psi\left(x\left(t \right) \right) = \int \frac{dp}{E_p} \psi_p \exp{i \left[E_p(t_0 + t)  - p (x_0 + v t) \right]}
\end{equation}
Without loss of generality we can consider $t_0=x_0=0$ (the phase given by different initial values can be reabsorbed in the definition of the state $\psi_p$), thus reducing the last expression to:
\begin{equation}
    \psi\left(x\left(t \right) \right) = \int \frac{dp}{E_p} \psi_p \exp{i \left[\frac{E_p}{v}  - p \right]v\, t}
\end{equation}
The best localization will be performed by a probe interacting with the particle for the shortest time possible, thus the probe velocity $v$ will be considered as approaching $c$:
\begin{equation}
\label{statedef}
    \psi\left(x\left(t \right) \right) \approx \int \frac{dp}{E_p} \psi_p \exp{i \left[\frac{E_p}{c}  - p \right]c\, t}
\end{equation}
As shown in \ref{newvar}, introducing the new variable $\xi = \frac{E_p}{c} - p$, the last integral can be written as:
\begin{equation}
    \psi\left(x\left(t \right) \right) = \int_0^{\infty} \frac{d\xi}{\xi} \psi_{p(\xi)}\, e^{i \xi\, c\, t}
\end{equation}
We thus arrive at the conclusion that the probability density to interact at a given instant $t$ on the world-line of a probe particle in a well-localized state is given by:
\begin{equation}
    \rho\left( t \right) \propto \left| \int_0^{\infty} \frac{d\xi}{\xi} \psi_{p(\xi)}\, e^{i \xi\, c\, t} \right|^2
\end{equation}

We can now proceed to show the existence of a minimal duration in the interaction. Using the probability density defined above we can measure the interaction duration as the square root of:
\begin{equation}
\label{intlength}
    \delta t^2 = \overline{t^2} \equiv \frac{\int dt \, t^2\,  \rho\left( x(t) \right) }{\int dt \,  \rho\left( x(t) \right)}
\end{equation}
where the average is given in terms of the probability density $ \rho\left( t \right)$,
and we performed a translation of the state in order to have $\overline{t} = 0$.
With a Fourier transform the last average can be expressed in $\xi$-space as:
\begin{equation}
\label{deltat}
    \delta t^2 = \frac{1}{c^4}\frac{\int_0^{\infty} d\xi \left| \frac{d}{d\xi} \frac{\psi_{p(\xi)}}{\xi}\right|^2 }{\int_0^{\infty} d\xi \left| \frac{\psi_{p(\xi)}}{\xi}\right|^2 } = \frac{ 1 }{c^4} \left\langle \left|\frac{d}{d\xi}\right|^2 \right\rangle_{\xi}
\end{equation}
This expression can be recognized as the uncertainty of the operator $\left(-i\frac{d}{d\xi}\right)$ evaluated on the state $\frac{\psi_{p(\xi)}}{\xi}$, with scalar product $\int_0^{\infty} d\xi$ (whose associated expectation value we denoted as $\left\langle\bullet \right\rangle_{\xi}$).
A bound can then be established on the value of  $\delta t^2$ using the uncertainty relation between $\left(-i\frac{d}{d\xi}\right)$ and the multiplicative operator $\xi$ on the Hilbert space defined by the scalar product\footnote{In order for $\left(-i\frac{d}{d\xi}\right)$ to be a self-adjoint operator, the states of this Hilbert space have to be functions of $\xi$ going to zero for $\xi \to 0$ and $\xi \to \infty$; this is guaranteed by the definition of the variable $\xi$, which links a fast decrease of states $\psi_p$ for $p \to \pm \infty$ to a fast decrease of $\psi_{p(\xi)}$ for  $\xi \to 0$ and $\xi \to \infty$.} $\int_0^{\infty} d\xi$.
We will show that the uncertainty in the variable $\xi$ has an upper bound on the class of states we are interested in, which implies a lower bound in the \rev interaction length" $\delta t^2$.


As usual, the uncertainty on $\xi$ and the one on $\left(-i\frac{d}{d\xi}\right)$ are inversely related. For the operators $x$ and $p$ in the standard Hilbert space the single uncertainties are not bounded, so even if one cannot know both of them arbitrarily well, we can choose a state very localized for, say, the $x$ operator at the cost of a diverging uncertainty for $p$. As we will see this is not the case in the relativistic theory, for which the uncertainty on the multiplicative operator $\xi$ is not allowed to be arbitrarily large (if the particle is on average at rest). From an upper bound in the $\xi$-uncertainty we will derive a lower bound  in the uncertainty of $\left(-i\frac{d}{d\xi}\right)$, which in turn will give a lower bound on the duration of any interaction with the particle we are trying to localize.
For this task it is useful to compute the first two momenta of $\xi$ on the state $\frac{\psi_{p(\xi)}}{\xi}$:
\begin{equation}
    \left\langle \xi \right\rangle_{\xi} = \frac{\int_0^{\infty} d\xi \, \xi \, \left| \frac{\psi_{p(\xi)}}{\xi}\right|^2 }{\int_0^{\infty} d\xi \left| \frac{\psi_{p(\xi)}}{\xi}\right|^2 }
\end{equation}

\begin{equation}
    \left\langle \xi^2 \right\rangle_{\xi} = \frac{\int_0^{\infty} d\xi \, \xi^2 \, \left| \frac{\psi_{p(\xi)}}{\xi}\right|^2 }{\int_0^{\infty} d\xi \left| \frac{\psi_{p(\xi)}}{\xi}\right|^2 }
\end{equation}
We have:
\begin{equation}
\label{bound1}
    \int_0^{\infty} d\xi \left| \frac{\psi_{p(\xi)}}{\xi}\right|^2 = \int \frac{dp}{E_p^2} \frac{  \left| \psi_{p}\right|^2  }{1 - v_p}
\end{equation}

\begin{equation}
\label{bound2}
    \int_0^{\infty} d\xi \, \xi \, \left| \frac{\psi_{p(\xi)}}{\xi}\right|^2 = \int \frac{dp}{E_p}   \left| \psi_{p}\right|^2
\end{equation}

\begin{equation}
\label{bound3}
    \int_0^{\infty} d\xi \, \xi^2 \, \left| \frac{\psi_{p(\xi)}}{\xi}\right|^2 = \int dp \, \left( 1- v_p \right)    \left| \psi_{p}\right|^2
\end{equation}
with $v_p = \frac{p}{E_p}$.

We are interested in establishing a bound in the interaction length with a particle at rest. If we translate this condition as a constraint on the average value of the momentum we can derive the condition for $\xi$:
\begin{equation}
    \int \frac{dp}{E_p}  \left| \psi_{p}\right|^2 p = \int_0^{\infty} \frac{d\xi}{\xi} \left| \psi_{p(\xi)}\right|^2  \frac{m^2 - \xi^2}{2 \xi} =0
\end{equation}
From the last equality it is easy to conclude that:
\begin{equation}
    \frac{\int_0^{\infty} d\xi \left| \psi_{p(\xi)}\right|^2}{\int_0^{\infty} d\xi\, \left|\frac{ \psi_{p(\xi)}}{\xi}\right|^2} = \left\langle \xi^2 \right \rangle_{\xi} = m^2
\end{equation}
in addition we have:
\begin{equation}
    \left\langle \xi \right\rangle_\xi = \frac{m^2}{\left\langle  E_p \right\rangle}
\end{equation}
where on the right-hand side we have the expectation value with the standard momentum measure $\int \frac{dp}{E_p}$. Combining the last two equations we obtain for $ \delta \xi^2$:
\begin{equation}
    \delta \xi^2 =  \left\langle \xi^2 \right \rangle_{\xi} - \left\langle \xi \right\rangle_\xi^2 = m^2 \left( 1 - \frac{m^2}{\left\langle  E_p \right\rangle^ 2} \right)
\end{equation}
This result gives us a constraint for the minimum duration of the interaction. The uncertainty principle between $\left(-i \frac{d}{d\xi}\right)$ and $\xi$ combined with \eqref{deltat} leads to:
\begin{equation}
\label{uncprin}
   \delta t^2  \geq \frac{1}{4 m^2  c^4 }  \frac{1}{ \left( 1 - \frac{m^2}{\left\langle  E_p \right\rangle^ 2} \right)}
\end{equation}
In this amount of time the probe covers a distance $v\, \delta t \approx c\, \delta t$, so we obtain an equivalent bound for the minimum spatial extension of the interaction:
\begin{equation}
\label{mindx}
    \delta x^2  \geq \frac{1}{4 m^2  c^2 }  \frac{1}{ \left( 1 - \frac{m^2}{\left\langle  E_p \right\rangle^ 2} \right)}
\end{equation}
We have to keep in mind that the coordinates $x$ and $t$ are linked to the probe, meaning that they indicate the spatial and temporal extension of the interaction \emph{on the probe world-line}. A bound imposed on them has an absolute value in that there is no probe capable of interacting for a shorter time with the particle, given the finiteness of the speed of light.

The inequality \eqref{mindx} poses an absolute limit on the localizability of the particle because it implies:
\begin{equation}
    \delta x^2  \geq \frac{1}{4 m^2  c^2 }
\end{equation}
which is the limit derived in \cite{landau1931erweiterung} in a heuristic way.
From \eqref{statedef} we see that in the galilean limit, $\frac{p}{m c} \to 0$, the $\xi$ coordinate is equivalent to $-p$, as the probe becomes infinitely fast and is capable of taking fixed-time snapshots of the particle state. With the relabeling $x = c\, t $ the state becomes, in this limit:
\begin{equation}
    \psi\left(x \right) = \int \frac{dp}{E_p} \psi_p \, e^{- i  p x}
\end{equation}
and the rescaled coordinate $x$ becomes the standard spatial coordinate.

Our main goal is to obtain the first special-relativistic correction to the Heisenberg principle that applies when $\delta p$ is small. This can be found by approximating \eqref{mindx} assuming we are close to the galilean limit. For this purpose we observe that
\begin{equation}
    \left\langle  \frac{E_p}{m c^2} \right\rangle^2 \simeq 1 + \frac{\delta p^2}{m^2 c^2} + \left(\frac{\delta p^2}{2 m^2 c^2}\right)^2 - \frac{\langle p^4 \rangle}{4 m^4 c^4}
\end{equation}
Assuming the minimum-uncertainty states to be continuous deformations of their galilean counterparts, we can exploit the zero-order identity $\langle p^4 \rangle = 3 \left( \delta p^2\right)^2$. Substituting in \eqref{mindx}, we find:
\begin{equation}
    \delta x^2 \delta p^2 \gtrsim \frac{1}{4} \left( 1 + \frac{3}{2}\frac{ \delta p^2 }{m^2 c^2} \right) 
\end{equation}
We feel that this is our main result. There may be some conceptual value also in the fact that our approach could be used to derive higher-order corrections to the Heinsenberg principle, but this leading-order special-relativistic correction is evidently our most robust result and also the result best suited for providing guidance to experimental tests of the interplay between special relativity and the Heisenberg principle, which, as stressed above, should naturally aim for reaching high accuracy at relatively small values of momentum uncertainty. We challenge experimentalists to achieve this goal. Future proposals of alternatives to our strategy of derivation of the special-relativistic corrections should be compared to ours mainly for what concerns the predictions for this leading-order correction and hopefully the superiority of one strategy over another will not provide material for endless \rev conceptual debates" but rather be established experimentally.

\section*{Acknowledgements}

G.A.-C.'s work on this project was supported by the FQXi grant 2018-190483 and by the MIUR, PRIN 2017 grant 20179ZF5KS.
We gratefully acknowledge valuable discussions with Giacomo Rosati and Giovanni Palmisano in the initial stages of this project.

\bibliography{riferimenti}

\begin{thebibliography}{10}
\expandafter\ifx\csname url\endcsname\relax
  \def\url#1{\texttt{#1}}\fi
\expandafter\ifx\csname urlprefix\endcsname\relax\def\urlprefix{URL }\fi
\expandafter\ifx\csname href\endcsname\relax
  \def\href#1#2{#2} \def\path#1{#1}\fi

\bibitem{amelino2013quantum}
G.~Amelino-Camelia, Quantum-spacetime phenomenology, Living Reviews in
  Relativity 16~(1) (2013) 1--137.

\bibitem{carlip2001quantum}
S.~Carlip, Quantum gravity: a progress report, Reports on progress in physics
  64~(8) (2001) 885.

\bibitem{landau1931erweiterung}
L.~Landau, R.~Peierls, Erweiterung des unbestimmtheitsprinzips f{\"u}r die
  relativistische quantentheorie, Zeitschrift f{\"u}r Physik 69~(1) (1931)
  56--69.

\bibitem{newton1949localized}
T.~D. Newton, E.~P. Wigner, Localized states for elementary systems, Reviews of
  Modern Physics 21~(3) (1949) 400.

\bibitem{freidel2007relativistic}
L.~Freidel, F.~Girelli, E.~R. Livine, Relativistic particle: Dirac observables
  and feynman propagator, Physical Review D 75~(10) (2007) 105016.

\bibitem{wigner1983interpretation}
E.~P. Wigner, Interpretation of quantum mechanics, in: Philosophical
  Reflections and Syntheses, Springer, 1983, pp. 78--132.

\bibitem{teller1994interpretive}
P.~Teller, An interpretive introduction to quantum field theory, Princeton
  University Press, 1994.

\bibitem{marolf2002relativistic}
D.~Marolf, C.~Rovelli, Relativistic quantum measurement, Physical Review D
  66~(2) (2002) 023510.

\bibitem{reisenberger2002spacetime}
M.~Reisenberger, C.~Rovelli, Spacetime states and covariant quantum theory,
  Physical Review D 65~(12) (2002) 125016.

\bibitem{itzykson2012quantum}
C.~Itzykson, J.-B. Zuber, Quantum field theory, Courier Corporation, 2012.

\end{thebibliography}
\appendix

\section{New variables}\label{newvar}
We define the light-cone variable in 3+1 dimensions as:
\[
\xi_p = E_p - p_{\parallel}
\]
where $p_{\parallel}$ is the projection of the particle momentum on the direction of motion of the probe.
From this definition it is straightforward to derive:
\[
p_{\parallel}\left(\xi, E_{\perp}^2  \right) = \frac{E_{\perp}^2 - \xi^2}{2 \xi}
\]
\[
p^2 \left(\xi\right) = \frac{1}{4}\left( \frac{E_{\perp}^2}{\xi} + \xi \right)^2 - m^2
\]
with $E_{\perp}^2 = m^2 + p_{\perp}^2$, and $p_{\perp}$ the momentum component orthogonal to the direction of motion of the probe. From the last equation we see that any spherical function of $p$ can only depend on the combination $\left( \xi +  \xi^{-1}E_{\perp}^2\right)$.
We have, in addition:
\[
   dp_{\perp}^2 d\xi = \left| \frac{\partial \xi_p}{\partial p_{\parallel}}\right| dp^3 = \left|\frac{p_{\parallel}}{E_p} - 1 \right| dp^3 \Longrightarrow \frac{dp_{\perp}^2 d\xi}{\xi} = \frac{dp^3}{E_p}
\]

In 1+1 dimensions these relations reduce to:
\[
    \xi_p = E_p - p
\]

\[
    p\left({\xi}\right) = \frac{m^2 - \xi^2}{2\xi}
\]

\[
    d\xi = \left| \frac{d\xi_p}{dp} \right| dp = \left|\frac{p}{E_p} - 1 \right| dp \Longrightarrow \frac{d\xi}{\xi} = \frac{dp}{E_p}
\]



\end{document}